\documentclass[preprint,showpacs,preprintnumbers,amsmath,amssymb,aps,pre]{revtex4-1}

\usepackage{natbib}
\usepackage{graphicx}
\usepackage{dcolumn}
\usepackage{bm}
\usepackage{subfigure}

\begin{document}


\title{Sets of FCS experiments to quantify free diffusion coefficients in reaction-diffusion systems. The case of $Ca^{\rm 2+}$and its dyes}

\author{Lorena Sigaut, Cecilia Villarruel, Mar\'\i a Laura Ponce and Silvina Ponce Dawson}
\affiliation{Departamento de
  F\'\i sica, FCEN-UBA, and IFIBA, CONICET,\\
    Ciudad Universitaria, Pabell\'on I, (1428) Buenos Aires, Argentina}


\begin{abstract}
Many cell signaling pathways involve the diffusion of
\textit{messengers} that bind/unbind to intracellular components.
Quantifying their net transport rate under different conditions, then
requires having separate estimates of their free diffusion coefficient
and binding/unbinding rates. In this paper, we show how performing sets
of \textit{Fluorescence Correlation Spectroscopy} (FCS) experiments
under different conditions, it is possible to quantify free diffusion
coefficients and on and off rates of reaction-diffusion systems. We
develop the theory and present a practical implementation for the case
of the universal second messenger, calcium (Ca$^{2+}$) and
single-wavelength dyes that increase their fluorescence upon Ca$^{2+}$
binding. We validate the approach with experiments performed in aqueous
solutions containing Ca$^{2+}$ and Fluo4 dextran (both in its High and
Low Affinity versions). Performing FCS experiments with
tetramethylrhodamine-dextran in \textit{Xenopus laevis} oocytes, we
infer the corresponding free diffusion coefficients in the cytosol of
these cells. Our approach can be extended to other physiologically
relevant reaction-diffusion systems to quantify biophysical parameters
that determine the dynamics of various variables of interest. 

\end{abstract}


\maketitle

\section{Introduction}

Many cell signaling pathways involve the diffusion of messengers in the
cytoplasm. In most cases these substances convey their message by
binding to target molecules. Furthermore, as they reach their targets
they not only diffuse freely but also bind/unbind to other cell
components. For long enough times, the net transport that results from
the combination of free diffusion and binding/unbinding is described by
\textit{effective diffusion coefficients} that are weighted averages of
the free coefficients of the messenger and of the substance it
interacts with that depend on their concentrations and reaction rates.
The universal second messenger calcium ($Ca^{\rm 2+}$) provides a
prototypical example of this behavior. Persistently high cytosolic
$Ca^{\rm 2+}$ concentrations lead to cell death. For this reason cells
have numerous mechanisms to reduce this concentration, the fastest one
of which is \textit{buffering}. Buffers are molecules that bind/unbind
to/from $Ca^{\rm 2+}$ ions reducing their free concentration. In doing
so they also alter the spatio-temporal cytosolic $Ca^{\rm 2+}$
distribution~\cite{Dargan2003},~\cite{Dargan2004} and the effect of the
resulting signals on the eventual end responses~\cite{Nelson1995}. This
means that the time and spatial range of action of $Ca^{\rm 2+}$ as
messenger is strongly dependent on the $Ca^{\rm 2+}$ concentration
itself. There is a large variety of intracellular $Ca^{\rm 2+}$ signals
ranging from those that arise upon the opening of a single $Ca^{\rm
2+}$ channel on the plasma membrane or on the membrane of intracellular
stores to those that manifest themselves as $Ca^{\rm 2+}$ waves that
travel throughout the whole cell~\cite{Sun1998}. This in turn implies
that the cytosolic $Ca^{\rm 2+}$ concentration attains very different
values depending on the signal type. The resulting $Ca^{\rm 2+}$
effective diffusion coefficient then varies across disparate values
depending on the signal type. The range of values was estimated to be
$\sim$ [5, 220] \textbf{$\mu m^{\rm 2} {\rm /}s$}~\cite{Allbritton1992}
with increasing values as the cytosolic $Ca^{\rm 2+}$ concentration
increased. How fast can $Ca^{\rm 2+}$ diffuse inside cells? In order to
answer this question it is necessary to have separate estimates of the
$Ca^{\rm 2+}$ and $Ca^{\rm 2+}$ buffers free diffusion coefficients,
their concentrations and reactions rates. Ideally, having access to
this information one could eventually compute the $Ca^{\rm 2+}$
effective diffusion coefficient as a function of its concentration. One
could wonder why having access to this information would be necessary
to study $Ca^{\rm 2+}$ signals. After all, they can be observed in
intact cells using $Ca^{\rm 2+}$ dyes. The $Ca^{\rm 2+}$ dye
fluorescence, however, provides information on the $Ca^{\rm 2+}$-bound
dye concentration which distribution depends on the dye kinetic and
transport properties. In particular, $Ca^{\rm 2+}$ signals that are
evoked via the photo-release of caged compounds with UV light are
imaged using single wavelength dyes that increase their fluorescence
when bound to $Ca^{\rm 2+}$~\cite{Paredes2008},~\cite{Piegari2014}. It
was shown in Bruno et al., 2010~\cite{Bruno2010} that $Ca^{\rm 2+}$
current estimates inferred from images that use such dyes are quite
sensitive to uncertainties in the on rate of the $Ca^{\rm 2+}$-dye
binding reaction and in the diffusion coefficient of the dye, two
parameters that are usually poorly known. Having reliable estimates of
these parameters is thus unavoidable to extract quantitative
information from images of $Ca^{\rm 2+}$ signals and allow the
effective interplay between modeling and experiment that is necessary
to attain a comprehensive description of the signals. In this paper we
describe and implement an approach that shows how performing sets of
\textit{Fluorescence Correlation Spectroscopy} (FCS)~\cite{Elson2011}
experiments under different conditions and using a reaction-diffusion
model to interpret the experimental results it is possible to obtain
separate estimates of these key biophysical parameters.

In FCS the fluorescence intensity in a small volume is recorded along
time and, via an analysis of the temporal autocorrelation of the
observed fluctuations, the transport rates of the fluorescent species
are, in principle, derived~\cite{Magde1972}. FCS has been widely used
to determine the diffusion coefficients of fluorescently labeled
proteins inside cells~\cite{Elson2001}, \cite{Schwille_2007},
\cite{Bismuto2001}. When the fluorescent species diffuse freely in 
mediumthere is an analytic expression for the autocorrelation function
of the fluorescent fluctuations (ACF) that is used to fit the
experimental observations and derive diffusion coefficients (see
Materials and Methods). When the fluorescent particles diffuse and
react, simple analytic expressions for the ACF can only be obtained
under certain approximations~\cite{Bismuto2001}, \cite{Ipina2013},
\cite{Ipina2014}, \cite{Sigaut2010}. In particular, we have shown in
Sigaut et al., 2010~\cite{Sigaut2010} (see also~\cite{Ipina2013},
\cite{Ipina2014}), that when the reactions occur on a somewhat faster
timescale than diffusion the correlation time of the fluctuations is
determined by the effective diffusion coefficients mentioned before.
Our theoretical studies showed that information on reaction rates could
also be derived from the fitting~\cite{Ipina2013},~\cite{Sigaut2010}.
In this paper we present a practical implementation of such an approach
in which the experimental conditions are changed so as to maximize the
information that can be drawn from the data. More specifically, we do
it for the case of {$Ca^{\rm 2+}$ }and single wavelength $Ca^{\rm 2+}$
dyes. This case shares some common features with the case in which
proteins diffuse and react, but it differs slightly since fluctuations
are also due to changes in fluorescence intensity associated to the
$Ca^{\rm 2+}$-dye binding/unbinding reaction. This requires the
development of a new theoretical framework that we introduce in this
paper as well.

In order to advance with the practical implementation presented here we
first study theoretically the behavior of the ACF for a case with
$Ca^{\rm 2+}$ and a single wavelength dye. We derive an analytic
approximation under the assumption that the $Ca^{\rm 2+}$-dye reaction
occurs on a fast timescale. We compare this approximated ACF with the
one without approximations computed numerically and determine the range
of parameter values for which the approximated ACF can give reasonable
estimates of certain parameters. We then show the results of a series
of FCS experiments performed in aqueous solutions containing $Ca^{\rm
2+}$ and different amounts of the $Ca^{\rm 2+}$ indicator Fluo4
dextran both in its High and Low Affinity versions
(Invitrogen-Molecular Probes, Carlsbad, CA). Fitting the observed ACF
by the analytic approximation we corroborate the validity of the
approximations and derive diffusion coefficients and the off-rate of
the $Ca^{\rm 2+}$-dye binding reaction, in solution. A similar
approach can be used to characterize the kinetic properties of other
$Ca^{{\rm 2+}}$ dyes. Even if the free diffusion coefficients in
solution and in the cytosol are different due to differences in
viscosity between both media, we may assume that the ratio between the
free diffusion coefficients of any two substances remains the same in
both settings. This is particularly relevant, because by solely
quantifying the rate of diffusion of a molecule that diffuses freely in
the cytosol and in solution we can infer the free diffusion coefficient
of $Ca^{\rm 2+}$and the dyes in the cytosol as well. We present such
quantification in the Appendix. Thus, the practical implementation
presented in this paper not only highlights the advantages of our
approach but it also allows us to derive information that is key to
quantify the free $Ca^{{\rm 2+}}$ distribution that underlies a
$Ca^{\rm 2+}$ image.

$Ca^{\rm 2+}$ signals are not the only example in which being able to
tell apart the contributions of free diffusion and reactions on the net
transport rate of labeled substances is relevant. We have recently
shown~\cite{Sigaut2014} the necessity of going beyond the description
of effective coefficients to interpret correctly the apparently
disparate estimates of the protein, Bicoid, diffusion coefficient
derived from FCS~\cite{AbuArish:BJ2010} and Fluorescence Recovery After
Photobleaching (FRAP)~\cite{Bialek:Cell2007} experiments. This example
also shows that the comprehensive quantifiable description of a
physiological process requires having a biophysical model for the
dynamics of the relevant concentrations that depends on
concentration-independent biophysical parameters. It is via such a
model that the response of the system over time in front of different
stimuli can be predicted. Being able to derive estimates of the
concentration-free biophysical parameters \textit{in situ} is thus
relevant to achieve a meaningful description. The approach presented in
this paper can be adapted and applied to other problems. Therefore, its
relevance goes beyond quantifying the biophysical parameters associated
to $Ca^{\rm 2+}$ and its dyes.

\section{Materials and Methods}

\subsection{FCS Theory}

Fluorescence Correlation Spectroscopy (FCS) monitors the fluctuation of
the fluorescence in a small volume. Fluctuations are characterized by
the time-averaged autocorrelation function (ACF) which is defined as:

\begin{equation}
\label{eq:ACF}
G(\tau )=\frac{\left\langle \delta f(t){\kern 1pt} \delta f(t+\tau )\right\rangle }{\left\langle f(t)\right\rangle ^{2} } 
\end{equation}

where $\langle f(t)\rangle$ is the average fluorescence in the sampling
volume and $\delta f(t)$ is the deviation with respect to this mean at
each time, $t$. As explained in the Appendix, when the fluorescence
comes from a single species, $P_{f}$, that diffuses freely with
coefficient $D_{f}$ (i.e., does not react) the ACF is of the form:

\begin{equation}
\label{eq:G1comp}
G(\tau )=\frac{Go}{\left({\rm 1}+\frac{\tau }{\tau _{f} } \right)\sqrt{{\rm 1}+ \frac{\tau }{w^{{\rm 2}} \tau _{f} } } } 
\end{equation}

where $w=w_{z} /w_{r}$ is the aspect ratio of the sampling volume and
$w_{z}$ and $w_{r}$ are the sizes of the beam waist along $z$ and $r$,
with $z$ the spatial coordinate along the beam propagation direction
and $r$ a radial coordinate in the perpendicular plane; the effective volume is $V_{ef} \equiv
\pi ^{3/2} w_{r}^{{\rm 2}} w_{z}$; $\tau _{f} =w_{r}^{{\rm 2}} /({\rm
4}D_{f} )$ is the characteristic time of diffusion of the particles
across the sampling volume and $Go=G(\tau ={\rm 0})={\rm 1/(}V_{ef}
P_{tot} )$, where $P_{tot}$ is the particle concentration. When the
dynamics of the fluorescent species is described by a
reaction-diffusion model most often there is not a simple analytic
expression for the ACF. It can always be written as a sum of integrals
each one associated to one of the branches of eigenvalues that rule the
dynamics of the linearized reaction--diffusion equations of the model.
Each of these integrals is called a ``component'' of the ACF. In the
case of interest for the present paper there are three relevant
species: free $Ca^{\rm 2+}$ (Ca), free dye ($F$) and $Ca^{\rm 2+}$-bound dye
($CaF$), which diffuse with free coefficients $D_{Ca}$ ($Ca$), 
and $D_{F}$ ($F$ and $CaF$) and react according to:

\begin{eqnarray}
\label{eq:reaction}
Ca + F\,\,\raisebox{-2.5ex}{$\stackrel{\stackrel{\textstyle k_{off}}
{\textstyle{\longleftarrow }}}{\stackrel{\textstyle{\longrightarrow}}
{\textstyle{k_{on}}}}$} \,\, CaF 
\end{eqnarray}

with on- and off-rates, $k_{on}$ and $k_{off}$ respectively. 
The corresponding (spatially uniform) equilibrium concentrations,
$Ca_{eq}$, $F_{eq}$ and $CaF_{eq}$ satisfy:

\begin{equation}
\label{eq:CaF_eq}
CaF_{eq} =\frac{Ca_{eq} F_{tot} }{Ca_{eq} +K_{d} } 
\end{equation}

where $K_{d} =k_{off} {\rm / }k_{on}$ and $F_{tot} =F_{eq} +CaF_{eq}$
is the total dye concentration. There are three branches of eigenvalues
for this system and the ACF then has three components. Simple algebraic
expressions can be obtained for the components in certain limits. In
particular, in this paper we present the results obtained in the ``fast
reaction limit'' that holds when the characteristic time of the
reaction Eq.~(\ref{eq:reaction}) is shorter than the time it takes for
the species to diffuse across the observation volume (i.e., if 
$\tau _{reac} \equiv \left(k_{off} +k_{on} (Ca_{eq} +F_{eq} )\right)^{-1} 
< w_{r} ^{2} /(4D_{F} )$). For more details see the Appendix where we also 
compare the ``full'' (integral expression of the) ACF computed numerically 
with the analytic approximation derived in the fast reaction limit that is 
presented in the Results Section and some of their components separately.

\subsection{FCS Experiments}

\subsubsection{Aqueous solutions}

Aqueous solutions were prepared with different concentrations of the
$Ca^{\rm 2+}$ indicator Fluo4 dextran High or Low Affinity (Invitrogen-
Molecular Probes, Carlsbad, CA), employing the solutions of a $Ca^{\rm
2+}$ Calibration Buffer Kit (Invitrogen- Molecular Probes). Each
solution contained 4.3 $\mu M$ $Ca^{\rm 2+}$, 100 mM KCl, 30 mM MOPS,
pH 7.2, and different concentrations of the $Ca^{\rm 2+}$ indicator
ranging from 200 $nM$ to 9$\mu M$ and from 400 $nM$ to 20$\mu M$ for
the Low and High Affinity version, respectively. Four or five separate
experiments were performed for each solution. Some of the results were
finally discarded as explained later. The solutions that were probed
and fitted are listed in Table \ref{tab:soluciones_F4_Ca}.

\begin{table}[!t]
\centering
\caption{\label{tab:soluciones_F4_Ca} FCS experiments in aqueous solutions containing Fluo4: Solutions composition.}
\begin{tabular}{c c c}
\\ \hline\hline
Aqueous solution & $Ca_{tot} (nM)$ & $F4_{tot} (nM)$\\ 
\hline
\multicolumn{3}{c}{Fluo4 High Affinity}\\ 
\hline
1 & 4285& 429\\ 
2 & 4285& 857\\     
3 & 4285& 1371\\   
4 & 4285& 1886\\ 
5 & 4285& 2571\\ 
6 & 4285& 4286\\ 
7 & 4285& 9000\\ 
8 & 4285& 15000\\ 
9 & 4285& 19286\\ 
\hline
\multicolumn{3}{c}{Fluo4 Low Affinity}\\ 
\hline
10 & 4285& 214\\ 
11 & 4285& 429\\ 
12 & 4285& 857\\ 
13 & 4285& 1114\\ 
14 & 4285& 1371\\ 
15 & 4285& 1886\\ 
16 & 4285& 2571\\ 
17 & 4285& 4286\\ 
18 & 4285& 9000\\ 
\hline
\multicolumn{3}{c}{All the solutions also contain:}\\
\multicolumn{3}{c}{ 100~mM~KCl, 30~mM~MOPS at pH~7.2 }\\
\hline\hline
\end{tabular}
\end{table}

\subsubsection{Acquisition}

FCS measurements were performed on a spectral confocal scanning
microscope FluoView 1000 (Olympus, Tokyo, Japan), employing a 60x, 1.35
N.A. oil-immersion objective (UPlanSAPO, Olympus) and a pinhole
aperture of 115~$\mu m$. Single point measurements at a 50 kHz sampling
rate were performed for a total duration of 167~s (equivalently, 8365312
data points) employing a 488~nm line and detecting the fluorescence in
the range (500-600)~nm. The measurements were performed at
$\sim$20~$\mu m$ from the coverslip.

\subsubsection{Data analysis}

Experimental ACF's were calculated with a custom-made routine written
on the Matlab platform~\cite{MATLAB}. To this end, each 167~s long
record was divided into $N=1021$, 164~ms long segments containing
2${}^{13}$ points each for the experiments in aqueous solutions. The
ACF was computed for each of the $N=1021$ segments from which the
average ACF was obtained. Based on the theoretical calculations
presented in the Results Section, we fitted the average ACF by an
expression of the form

\begin{equation}
\label{eq:G3comp}
G(\tau )=\frac{G_{0} }{\left({\rm 1}+\frac{\tau }{\tau _{0} } \right)\sqrt{{\rm 1}+\frac{\tau }{w^{{\rm 2}} \tau _{0} } } } +
\frac{G_{1} }{\left({\rm 1}+\frac{\tau }{\tau _{1} } \right)\sqrt{{\rm 1}+\frac{\tau }{w^{{\rm 2}} \tau _{{\rm 1}} } } } + 
\frac{G_{2} e^{-\nu \tau } }{\left({\rm 1}+\frac{\tau }{\tau _{2} } \right)\sqrt{{\rm 1}+\frac{\tau }{w^{{\rm 2}} \tau _{2} } } }  
\end{equation}

where $w=w_{z} /w_{r}$ is the aspect ratio of the sampling volume, as
before, and the various times are related to diffusion coefficients by
~$\tau _{i} =w_{r}^{{\rm 2}} /({\rm 4}D_{i} )$, $i=1,2,3$, with the
beam waist, $w_{r}$. Only experiments for which the mean fluorescence
in the observation volume remained approximately constant during the
whole record were fitted. Experiments for which the average ACF was too
noisy were also discarded. In all cases we tried to fit the experiments
leaving all 7 parameters of Eq.~(\ref{eq:G3comp})~($G_{0}$, $G_{1}$,
$G_{2}$, $\tau _{0}$, $\tau _{1}$, $\tau _{2}$, $\nu$) free to be
fitted. In others we set $G_{2} =0$ and only derived $G_{0}$, $G_{1}$,
$\tau _{0}$ and $\tau _{1}$. Thus, we tried a 3-component and a
2-component fit for each experiment. All fitting parameters were
determined for each average ACF via a nonlinear least squares fit using
the Matlab built-in function nlinfit. In the figures we show the
average of the displayed fitting parameters and the average error
computed over all the experiments in a given set.

\subsubsection{Characterization of the confocal volume}

The radial beam waist and the aspect ratio were determined to be $w_{r}
{\rm =}$ 0.262 - 0.292 $\mu m$ and $w=w_{z} {\rm /}w_{r} {\rm =\; 5}$
by measuring the translational three-dimensional diffusion of
fluorescein (Sigma, St. Louis, MO) in buffer solution pH 9, assuming a
diffusion coefficient of 425 $\mu m^{{\rm 2}} {\rm
/}s$~\cite{Culbertson2002}. Thus, the resulting effective volume was
$V_{ef} {\rm =}$ (0.59 ${\rm \pm }$ 0.1) $\mu m^{{\rm 3}}$.

\section{Results}

\subsection{ FCS theory for a solution with $Ca^{\rm 2+}$ and a single wavelength dye in the limit of fast reactions}

Proceeding as described in the Appendix we determine that, in the fast
reaction limit for the case of a solution of $Ca^{{\rm 2+}}$ and a dye
the ACF of the fluorescence fluctuations can be approximated by the sum
of three components of the form:

\begin{equation}
\label{eq:Gapprox}
G_{approx} (\tau ){\rm =}G_{F} (\tau )+G_{ef{\rm 1}} (\tau )+G_{ef{\rm 2}} (\tau )
\end{equation}

\begin{equation}
\label{eq:GF}
G_{F} (\tau )=\frac{Go_{F} }{\left({\rm 1}+\frac{\tau }{\tau _{F} } \right)\sqrt{{\rm 1}+\frac{\tau }{w^{{\rm 2}} \tau _{F} } } }
\end{equation}

\begin{equation}
\label{eq:Gef1}
G_{ef1} (\tau )=\frac{Go_{ef{\rm 1}} }{\left({\rm 1}+\frac{\tau }{\tau _{ef{\rm 1}} } \right)\sqrt{{\rm 1}+\frac{\tau }{w^{{\rm 2}} \tau _{ef{\rm 1}} } } }  \end{equation}

\begin{equation}
\label{eq:Gef2}
G_{ef2} (\tau )=\frac{Go_{ef{\rm 2}} e^{-\nu _{F} \tau } }{\left({\rm 1}+\frac{\tau }{\tau _{ef{\rm 2}} } \right)\sqrt{{\rm 1}+\frac{\tau }{w^{{\rm 2}} \tau _{ef{\rm 2}} } } } 
\end{equation}

where $\tau _{F} =w_{r}^{{\rm 2}} /({\rm 4}D_{F} )$ and $\tau _{efi}
=w_{r}^{{\rm 2}} /({\rm 4}D_{efi} )$, $i={\rm 1},{\rm 2}$, with: 

\begin{equation}
\label{eq:Def1}
D_{ef{\rm 1}} =\frac{D_{Ca} +\alpha D_{F} }{{\rm 1}+\alpha } 
\end{equation}

\begin{equation}
\label{eq:Def2}
D_{ef{\rm 2}} =\frac{\alpha D_{Ca} +D_{F} }{{\rm 1}+\alpha }  
\end{equation}

and $\alpha =\frac{F_{eq}^{{\rm 2}} }{F_{tot} K_{d} }$ and: 

\begin{equation}
\label{eq:nu}
\nu _{F} =k_{off} +k_{on} (F_{eq} +Ca_{eq} )
\end{equation}

The weights $Go_{F}$ , $Go_{ef{\rm 1}}$ and $Go_{ef{\rm 2}}$ are given
by: 

\begin{equation}
\label{eq:GoF}
Go_{F} =\frac{{\rm 1}}{V_{ef} F_{tot} } 
\end{equation}

\begin{equation}
\label{eq:Goef1}
Go_{ef{\rm 1}} =\frac{{\rm 1}}{V_{ef} CaF_{eq} } \frac{F_{eq}^{{\rm 2}} }{F_{tot} K_{d} } \frac{K_{d} }{(K_{d} +F_{eq} +Ca_{eq} )} 
\end{equation}

\begin{equation}
\label{eq:Goef2}
Go_{ef{\rm 2}} =\frac{{\rm 1}}{V_{ef} CaF_{eq} } \frac{K_{d} }{(K_{d} +F_{eq} +Ca_{eq} )}  
\end{equation}

The sum of all the weights is inversely proportional to the concentration of fluorescent particles: 

\begin{equation}
\label{eq:Gotot}
Go_{tot} =Go_{ef{\rm 1}} +Go_{ef{\rm 2}} +Go_{F} =\frac{{\rm 1}}{V_{ef} CaF_{eq} }
\end{equation}

The sum of the two effective diffusion coefficients satisfies:

\begin{equation}
\label{eq:Def1+Def2}
D_{ef{\rm 1}} +D_{ef{\rm 2}} =D_{Ca} +D_{F}   
\end{equation}

As in Sigaut et al., 2010~\cite{Sigaut2010}, the approximate analytic
expression of the ACF given by Eqs.~(\ref{eq:Gapprox})-(\ref{eq:Gef2})~
is always valid for large enough $\tau$. The first term, however, is
exact. Thus, we can expect to be able to derive $D_{F}$ from all FCS
experiments. The approximations of $Go_{ef{\rm 1}}$ and $Go_{ef{\rm
2}}$ are valid for the values of $\tau$ that are relevant to determine
$\tau _{ef1}$ and $\tau _{ef2}$ from a fit to the experiments if $\tau
_{reac} \equiv \left(k_{off} +k_{on} (Ca_{eq} +F_{eq})\right)^{-1}
<<w_{r} ^{2} /(4D_{Ca})$.

\subsection {Fitting parameters from FCS experiments in aqueous solutions with $Ca^{\rm 2+}$and Fluo4 dextran}

\begin {figure} [!b]
\begin {center}
\includegraphics {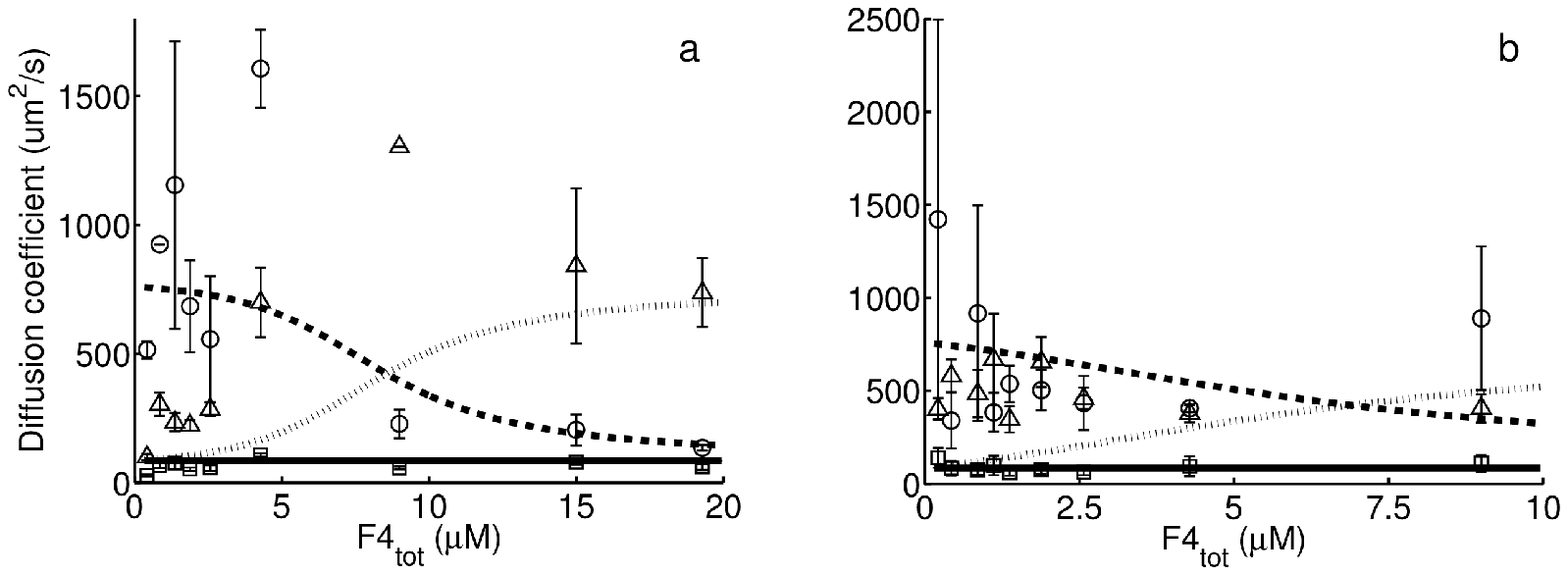} 
\end {center}
\caption {\label{fig:D_efectivos}}~Diffusion coefficients obtained from
the fitting of the experimental data using Eq.~(\ref{eq:G3comp}),
$D_{0}$ (squares), $D_{\rm 1}$(circles), and $D_{\rm 2}$(triangles), as
a function of the total calcium dye concentration of the aqueous
solutions, $F4_{tot}$. In solid line, $D_{F} = 85~\mu m^{2} /s$, and in
dash lines expected effective diffusion coefficients, $D_{ef{\rm 1}}$
(bold) and $D_{ef{\rm 2}}$ (light) given by Eqs.~(\ref{eq:Def1}) and
(\ref{eq:Def2}) respectively, with the calcium and dye concentrations
employed in the aqueous solutions, $D_{Ca} {\rm =}$ 760 $\mu m^{{\rm
2}} {\rm /}s$, $D_{F} {\rm =}$ 85 $\mu m^{{\rm 2}} {\rm /}s$ and the
dissociation constant given by the manufacturer, $K_{d} {\rm =}$ 772 nM
and 2600 nM for High (a) and Low (b) Affinity Fluo4.
\end {figure}

In this Section we show how we proceed to analyze the experimental
data. In particular, we show the results of using Eq.~(\ref{eq:G3comp})
to fit the ACF's obtained from the set of experiments of Table
\ref{tab:soluciones_F4_Ca} performed with Fluo4 High and Low Affinity.
The fitting parameters are $G_{0}$, $G_{\rm 1}$, $G_{\rm 2}$, $\nu$ and
the characteristic times $\tau _{0}$, $\tau _{\rm 1}$ and $\tau _{\rm
2}$ from which we derive three diffusion coefficients $D_{0}$, $D_{\rm
1}$ and $D_{\rm 2}$ as explained before. Fig.~\ref{fig:D_efectivos}
shows the diffusion coefficients obtained in this way as a function of
the total concentration of the dye used in the solutions, $F4_{tot}$,
for High Affinity (Fig.~\ref{fig:D_efectivos}~(a)) and Low Affinity
(Fig.~\ref{fig:D_efectivos}~(b)) Fluo4. We also plot in these figures
the expected free diffusion coefficient of the dye, $D_{F} {\rm =}$85
$\mu m^{{\rm 2}} {\rm /}s$~\cite{Gennerich_Schild2002}, and effective
diffusion coefficients, $D_{ef{\rm 1}}$ and $D_{ef{\rm 2}}$, calculated
using Eqs.~(\ref{eq:Def1})~and~(\ref{eq:Def2}), with the dissociation
constant given by the manufacturer ($K_{d} {\rm =}$ 772~nM for High
Affinity and $K_{d} {\rm =}$2600~nM for Low Affinity), $D_{Ca} {\rm
=}$760 $\mu m^{{\rm 2}} {\rm /}s$~\cite{Qin1991}, $D_{F} {\rm =}$85
$\mu m^{{\rm 2}} {\rm /}s$ and the total calcium and dye concentrations
employed in the solutions.

The identification between the fitting parameters $G_{0}$, $D_{0}$,
$G_{\rm 1}$, $D_{\rm 1}$, $G_{\rm 2}$,$D_{\rm 2}$, $\nu$, and the seven
quantities, $Go_{F}$, $D_{F}$, $Go_{ef{\rm 1}}$, $D_{ef{\rm 1}}$,
$Go_{ef{\rm 2}}$, $D_{ef{\rm 2}}$, $\nu _{F}$, of the theoretical
formulas Eqs.~(\ref{eq:Gapprox})~-~(\ref{eq:Gef2}) is immediate in the
case of the last three which correspond to the only component with an
exponentially decaying term. For the other quantities it is not
difficult to make the correspondence because $D_{F} <D_{Ca}$ implies
that $D_{F} \le D_{ef1}$. Furthermore, as may be observed in
Fig.~\ref{fig:D_efectivos}, there is one diffusion coefficient obtained
from the fitting, $D_{0}$, that remains approximately invariant for all
the analyzed concentrations. This should correspond to the free
diffusion of the dye, $D_{F}$, which is concentration independent. In
this way we determine that $D_{F} =(65\pm 7)~\mu m^{{\rm 2}} /s$ in the
case of Fluo4 High Affinity and $D_{F} =(89\pm 8)~\mu m^{{\rm 2}} /s$ in
the case of Fluo4 Low Affinity. This is clearer in
Fig.~\ref{fig:D_efectivos}~(a), and the lowest constant diffusion
coefficient can also be identified in Fig.~\ref{fig:D_efectivos}~(b).
The other two diffusion coefficients obtained from the fitting, $D_{\rm
1}$ and $D_{\rm 2}$, change with the dye concentration. This means that
they are effective diffusion coefficients. Making the identifications
$D_{1} =D_{ef{\rm 1}}$ and $D_{2} =D_{ef{\rm 2}}$ we know that their
lower and upper limits are the free diffusion coefficients of the dye,
$D_{F}$, and of calcium, $D_{Ca}$, respectively. In fact, both $D_{\rm
1}$ and $D_{\rm 2}$ are larger than $D_{0}$. Furthermore, in
Fig.~\ref{fig:D_efectivos}~(a), $D_{\rm 1}$ decreases with $F4_{tot}$
while $D_{\rm 2}$ increases similarly to their theoretical
counterparts, $D_{ef{\rm 1}}$ and $D_{ef{\rm 2}}$. This shows the
validity of the identification between fitting and model parameters. A
similar trend can be observed in Fig.~\ref{fig:D_efectivos}~(b)
although not as clear as in Fig.~\ref{fig:D_efectivos}~(a). In any
case, we do make the identification $D_1=D_{ef1}$ and $D_2=D_{ef2}$
also in this case. It is remarkable that the obtained results seem
reasonable even outside the range of validity of the fast reaction
approximation.

\begin{figure}[!ht]
 \includegraphics {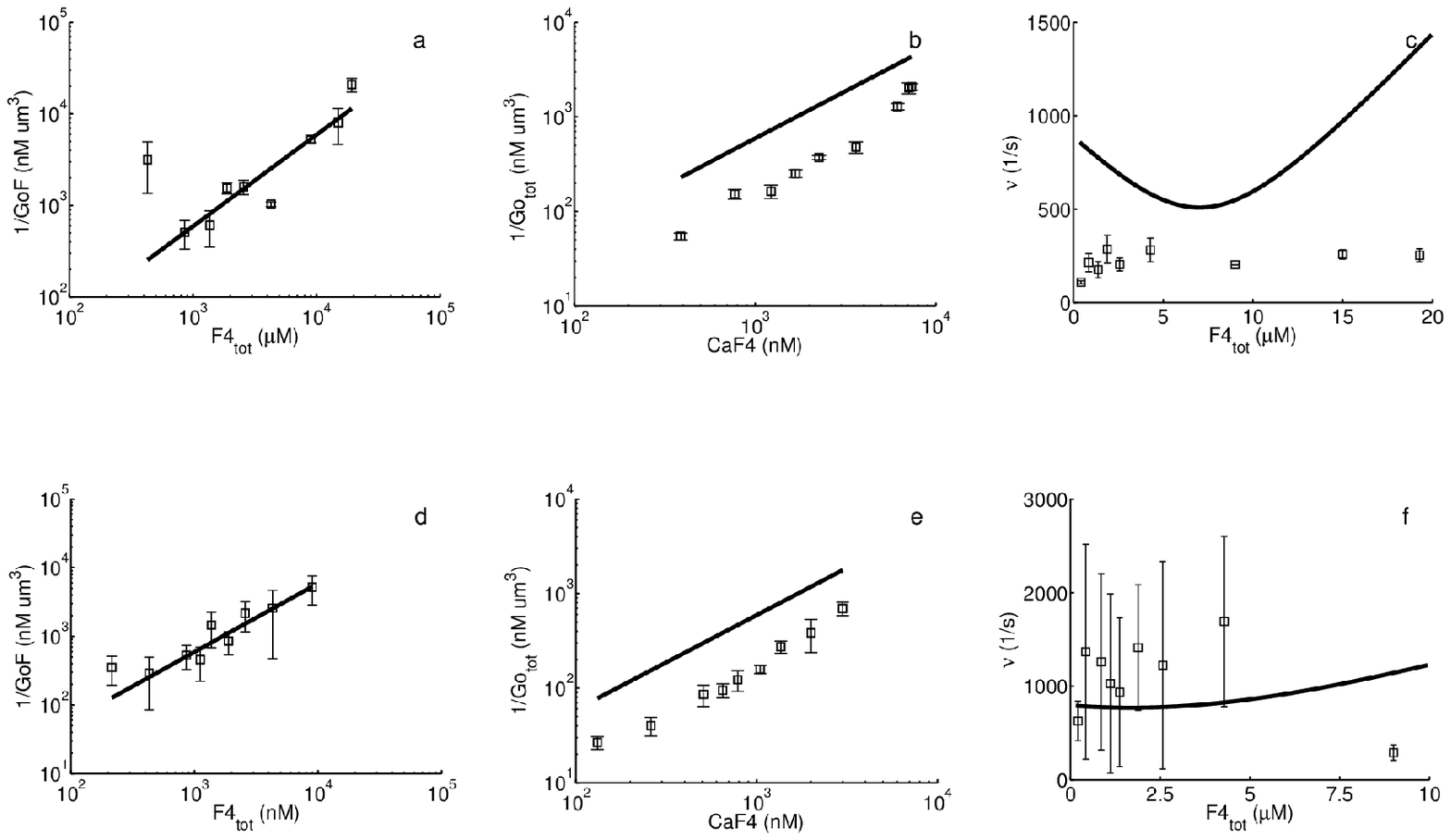} 
 \caption {\label{fig:GoF_Gotot_nu}} ~Parameters derived from the
 fitting of the experimental ACFs (with symbols) and theoretical
 expected values (solid lines). In a logarithmic scale, (left) inverse
 of the $Go_{F}$ as function of the total dye concentration used in the
 solutions, $F4_{tot}$, (middle) inverse of the sum of all the weights,
 $Go_{tot}$, as function of the $Ca^{{\rm 2+}}$-bound dye
 concentration, $CaF_{eq}$. $CaF_{eq}$ was estimated theoretically from
 Eq.~(\ref{eq:GoF}) with the $Ca^{\rm 2+}$ and dye concentrations of
 the aqueous solutions and $K_{d}$ given by the manufacturer. (Right)
 $\nu$ as function of the total dye concentration used in the
 solutions, $F4_{tot}$. (a, b, c) Fluo4 High Affinity and (d, e, f)
 Fluo4 Low Affinity.
\end{figure}

We test the validity of the theoretical approximation
Eqs.~(\ref{eq:GoF}) and (\ref{eq:Gotot})~in
Fig.~\ref{fig:GoF_Gotot_nu}. Fig.~\ref{fig:GoF_Gotot_nu} shows plots of
the inverse of $Go_{F}$ as a function of the total dye concentration
used in the solutions, $F4_{tot}$, with symbols, together with the
expected values given by Eq.(\ref{eq:GoF}) using the concentrations
used in the solutions and the observation volume derived from the
calibration, with curves. The results for High and Low Affinity Fluo4
are shown in Figs.~\ref{fig:GoF_Gotot_nu}~(a)
and~\ref{fig:GoF_Gotot_nu}~(d), respectively. The logarithmic scale
used in the figures highlights the fact that both the experimental and
the theoretical results scale similarly with $F4_{tot}$, i.e., as
1/$F4_{tot}$ (see Eq.~(\ref{eq:GoF})). If we fit the experimental
results using the values, $F4_{tot}$, determined by construction of the
solution, the effective volume, $V_{ef}$, can be obtained from the
fitting. Considering the inverse of $Go_{F}$ versus $F4_{tot}$, we
found expected values ($V_{ef} {\rm =}$ (0.54 ${\rm \pm }$ 0.08) $\mu
m^{{\rm 3}}$ and $V_{ef} {\rm =}$ (0.56 ${\rm \pm }$ 0.08) $\mu m^{{\rm
3}}$ for High Affinity and Low Affinity Fluo4, respectively) that are
consistent with the one obtained from the calibration ($V_{ef} {\rm =}$
(0.59 ${\rm \pm }$ 0.1) $\mu m^{{\rm 3}}$). 

Another property of the ACF is that the sum of all the weights,
$Go_{tot}$, is inversely proportional to the concentration of $Ca^{\rm
2+}$-bound dye, $CaF_{eq}$ (Eq.(\ref{eq:Gotot})). In
Figs.~\ref{fig:GoF_Gotot_nu}~(b) and~\ref{fig:GoF_Gotot_nu}~(e) we show
plots of the values of the inverse of $Go_{tot}$ obtained from the
fitting of the ACF as functions of $CaF_{eq}$ and the expected values
according to Eq.~(\ref{eq:Gotot}), for High Affinity and Low Affinity
Fluo4, respectively. The linear scaling between both quantities is very
good also in this case but there is a mismatch in the ordinate.
As before, we can fit the experimental results using the equilibrium
values, $CaF_{eq}$, derived from the concentrations used in the
solutions and the dissociation constant provided by the vendor.
Considering the inverse of $Go_{tot}$ versus $CaF_{eq}$ and fitting
with a linear relation, the effective volume inferred was (0.23 ${\rm
\pm }$ 0.02) $\mu m^{{\rm 3}}$ for Fluo4 High Affinity and (0.17 ${\rm
\pm }$ 0.01) $\mu m^{{\rm 3}}$ for the Low Affinity version of the dye
, which are lower than the one obtained from the calibration ($V_{ef}
{\rm =}$ (0.59 ${\rm \pm }$ 0.1) $\mu m^{{\rm 3}}$).

Finally we show the values of $\nu _{F}$ derived from the fitting and
the theoretical curve obtained using the fast reaction approximation,
Eqs.~(\ref{eq:Gapprox})-(\ref{eq:Gef2}), as a function of $F4_{tot}$
for High Affinity (Fig.~\ref{fig:GoF_Gotot_nu}~(c)) and Low Affinity
(Fig.~\ref{fig:GoF_Gotot_nu}~(f)) Fluo4. There we observe that the
values obtained for low $F4_{tot}$ concentrations are the ones that can
be associated to the theoretical expression (Eq. (\ref{eq:nu})) from
which an estimate of $k_{off}$ can be derived. In order to estimate
$k_{off}$ however we used all the data available as explained in the
Discussion. 

\subsection{Using the theory to determine free diffusion coefficients and reaction rates from the fitting}

\begin{figure}[!b]
\includegraphics {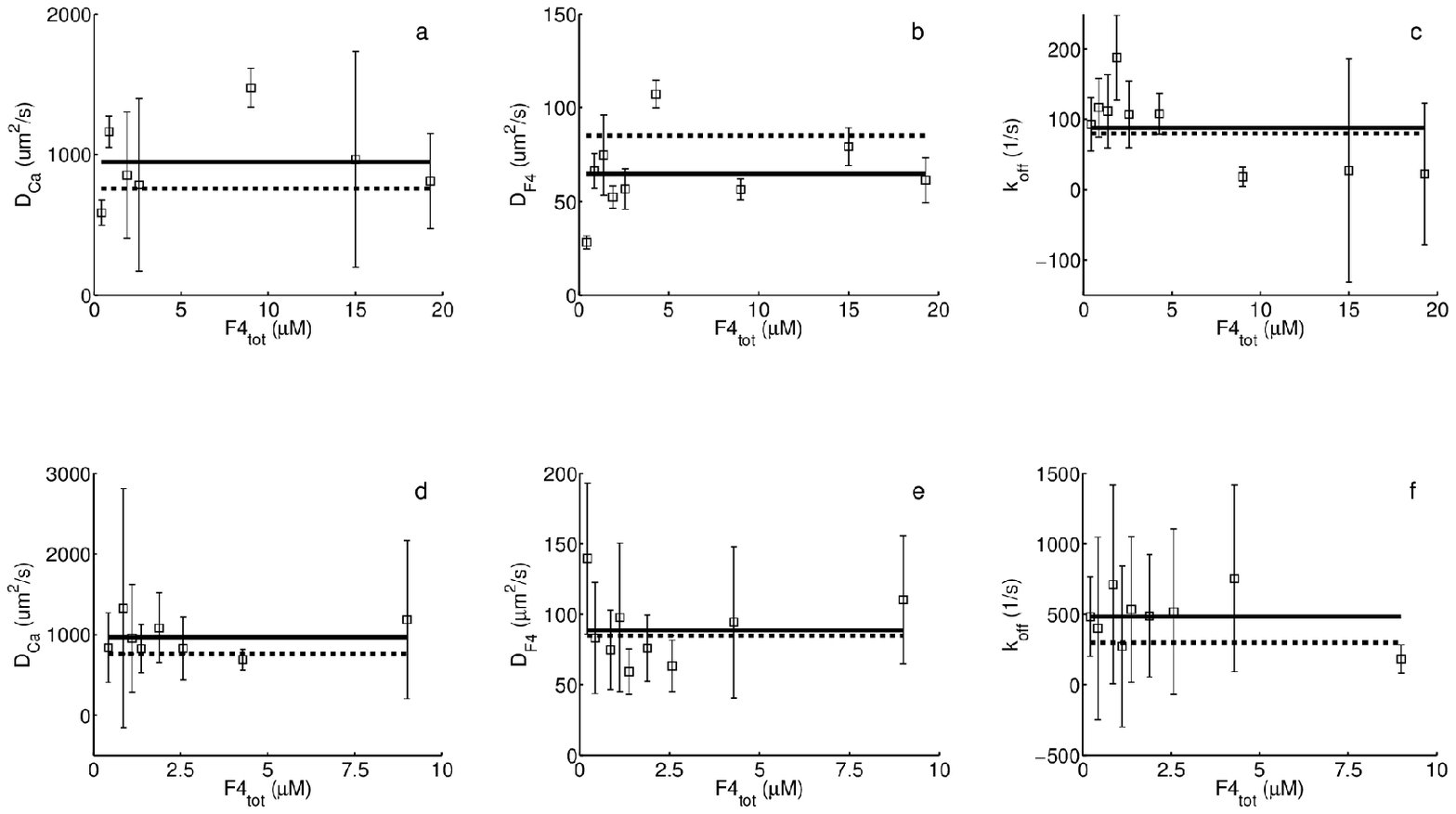}
\caption {\label{fig:DCa_DF_koff}} ~Parameters of the underlying
biophysical model derived from the fitting parameters for each aqueous
solution, $D_{Ca}$, $D_{F}$ and $k_{off}$, (mean and standard deviation
over 2-3 experiments with 1or 2 fits) and average of the values
obtained (solid line). (a, b, c) Fluo4 High Affinity and (d, e, f)
Fluo4 Low Affinity. In all cases we include the expected values (dashed
line) based on the total concentrations used in the solutions and on
previous estimates (see text).
\end{figure}

Being able to identify the parameters of the fitting with those of the
theoretical ACF, Eqs.~(\ref{eq:Gapprox})-(\ref{eq:Gef2}), allowed us to
go further and to quantify some relevant parameters of the underlying
biophysical model for each aqueous solution, such as the free calcium
diffusion coefficient. This entails solving an over-determined problem
(7 equations with 6 unknowns). In that sense, we preferred to use the
information given by $Go_{F}$ and $Go_{tot}$ rather than by $Go_{ef{\rm
1}}$ and $Go_{ef{\rm 2}}$ because, as discussed before, these weights
carry the largest errors. In particular, knowing $D_{F}$, $D_{ef{\rm
1}}$, $D_{ef{\rm 2}}$, $Go_{F}$, $Go_{tot}$ and $\nu _{F}$ (which we
identify with the 7 parameters of the fitting) it is possible to infer
the off-rate, $k_{off}$, of the $Ca^{\rm 2+}$-dye binding reaction, the
total concentration of the calcium dye, $F_{tot}$, the calcium bound
dye concentration in equilibrium, $CaF_{eq}$, and the free diffusion
coefficients, $D_{Ca}$, $D_{F}$. We show in Fig.~\ref{fig:DCa_DF_koff}
the values of $D_{Ca}$, $D_{F}$ and $k_{off}$ obtained as a function of
the total dye concentration used in the aqueous solutions, $F4_{tot}$,
both for the High Affinity (Figs.~\ref{fig:DCa_DF_koff}~(a)~-~(c)) and
the Low Affinity (Figs.~\ref{fig:DCa_DF_koff}~(d)~-~(f)) versions of
the dye. Since the solutions only differed in the total amount of dye,
all estimated parameter values, with the exception of $F4_{tot}$,
should remain approximately constant for all solutions. To estimate the
free $Ca^{\rm 2+}$ diffusion coefficient the solutions with effective
coefficients with large errors or with $D_{Ca}$ far away from the
average were discarded (solutions 3, 6, and 10). For Fluo4 High
Affinity we obtained $D_{Ca}$= (948 ${\rm \pm}$ 110) $\mu m^{\rm 2}/s$,
and if we also discard solution 7 which has also a $D_{Ca}$ that is
very different from the average, it gives $D_{Ca}$= (861 ${\rm \pm}$ 79)
$\mu m^{\rm 2}/s$. For Fluo4 Low Affinity we obtained $D_{Ca}$= (966
${\rm \pm}$ 76) $\mu m^{\rm 2}/s$, and if we also discard solutions 12
and 18, that have large errors, we obtained $D_{Ca}$= (870 ${\rm \pm}$
55)  $\mu m^{\rm 2}/s$. The average and standard deviation of all
estimated biophysical parameters are presented
in Table~\ref{tab:parametros}.

\begin{table}[!ht]
\centering
\caption {\label{tab:parametros} Reaction diffusion coefficients
estimated from the model. The results are expressed as mean $\pm$ SD.}
\begin{tabular}{c c c}
\hline\hline
Parameter & Estimation from the model & Previous estimates \\ 
\hline  
\multicolumn{3}{c}{High Affinity Fluo4} \\ 
\hline 
$D_{F}$ & (65 ${\rm \pm }$ 7) $\mu m^{{\rm 2}} {\rm /}s$ & 85 $\mu m^{{\rm 2}} {\rm /}s$~\cite{Gennerich_Schild2002} \\ 
$D_{Ca}$  & (948 ${\rm \pm }$ 110) $\mu m^{{\rm 2}} {\rm /}s$  & 760 $\mu m^{{\rm 2}} {\rm /}s$~\cite{Qin1991} \\ 
$k_{off}$ & (88 ${\rm \pm }$ 19) $1/s$ &\\
\hline 
\multicolumn{3}{c}{Low Affinity Fluo4} \\ 
\hline 
$D_{F}$  &  (89 ${\rm \pm }$ 8) $\mu m^{{\rm 2}}/s$ & 85 $\mu m^{{\rm 2}} {\rm /}s$~\cite{Gennerich_Schild2002}\\ 
$D_{Ca}$ & (966 ${\rm \pm }$ 76) $\mu m^{{\rm 2}} {\rm /}s$ &  760$\mu m^{{\rm 2}} {\rm /}s$~\cite{Qin1991} \\ 
$k_{off}$  & (483${\rm \pm }$ 61) $1/s$ &\\
\hline \hline 
\end{tabular}
\end{table}

\section{Discussion and conclusions}

In this work we have shown how free diffusion coefficients and reaction
rates can be quantified in reaction-diffusion systems by performing
sets of Fluorescence Correlation Spectroscopy (FCS) experiments and
using a biophysical model to interpret the experimental results. In
particular, we have applied this approach to the case of Ca$^{2+}$ and
a single wavelength Ca$^{2+}$ dye. To this end we developed the theory
that allowed us to derive an approximation of the autocorrelation
function of the fluorescence fluctuations (ACF) in the limit of fast
reactions. We then performed a series of experiments in solutions
containing  Ca$^{2+}$ and the  Ca$^{2+}$ dye Fluo4 dextran (both  High
and Low Affinity) with which validated the approach and established its
limitations. The analysis of the experiments also allowed us to
quantify the transport and reaction properties of two single wavelength
Ca$^{2+}$ dyes: High and Low Affinity Fluo4. In doing so we also
derived the free diffusion coefficient of Ca$^{2+}$ in aqueous
solution. Although this value is already well known ($D_{Ca}$ $\sim$
(750-800)~$\mu m^{{\rm 2}}/s$ ~\cite{Qin1991},~\cite{HCP}), being able
to derive it from the observation of a system in which it is not
diffusing freely is quite relevant and provides hints on how to proceed
in other settings.

Addressing fundamental problems in $Ca^{\rm 2+}$ signaling and
$Ca^{\rm 2+}$-dependent cell function calls for the use of multiple
approaches. The undeniable need to combine experiments and modeling
requires that key biophysical parameters such as the $Ca^{\rm 2+}$
diffusion coefficient be quantified \textit{in situ~}~\cite{von2014}.
Optical techniques are ideal to probe intracellular transport with
minimum disruption~\cite{Wachsmuth2015}. Measuring intracellular
$Ca^{\rm 2+}$ transport in this way, however, is not straightforward
because of the multiple interactions of the ions with different cell
components~\cite{Biess2011},~\cite{Bressloff2013} and because $Ca^{\rm
2+}$ dyes are also $Ca^{\rm 2+}$ buffers that alter the ions transport
rate~\cite{Piegari2015}. The quantification of diffusion coefficients
and reaction constants in such a case requires a careful
interpretation of the experimental data in terms of an underlying
biophysical model~\cite{Sigaut2014}. The work contained in this paper
constitutes a necessary first step to advance in this direction.

In this paper, we first focused on the theoretical aspects of the
problem. To this end, we derived an analytic approximation for the ACF
of a system with $Ca^{{\rm 2+}}$ and a single wavelength dye under the
assumption that the $Ca^{\rm 2+}$-dye reaction occurs on a fast
timescale, that the free and $Ca^{\rm 2+}$-bound dye molecules
diffuse at the same rate and that the former is not fluorescent. The
expression obtained, Eqs.~(\ref{eq:Gapprox})-(\ref{eq:Gef2}),
coincides with the one derived in Bismuto et al.,
2001~\cite{Bismuto2001}. In particular, we observed that the first two
terms in Eq.~(\ref{eq:Gapprox}) have the same functional dependence
on $\tau$ as the only term of Eq.~(\ref{eq:G1comp}) which corresponds
to a case with a purely diffusive species. The first term
(Eq.~(\ref{eq:GF})) gives $\tau _{F} =w_{r}^{\rm 2} {\rm /(4}D_{F}
)$ that corresponds to the dye diffusion time across the sampling
volume. This term is exact and involves no approximation. The second
term (Eq.~(\ref{eq:Gef1})) has the time scale $\tau _{ef{\rm 1}}
=w_{r}^{\rm 2} /({\rm 4}D_{ef{\rm 1}} )$ and is associated to an
effective diffusion coefficient, $D_{ef{\rm 1}}$, that combines
information on diffusion and reactions. $D_{ef{\rm 1}}$ corresponds to
the ``collective'' diffusion coefficient of Pando et al.,
2006~\cite{Pando2006} which in turn coincides with the effective
coefficient determined in the rapid buffering
approximation~\cite{Smith1996}. The last term 
(Eq.~(\ref{eq:Gef2})) does not have the functional form of a purely
diffusive case, but has an additional exponential factor. Depending on
the value of $\nu _{F}$, it could be neglected to determine $\tau
_{F}$ and $\tau _{ef{\rm 1}}$. In the Appendix we presented the
results of a thorough analysis of the limitations of this
approximation. In particular we computed numerically the ``full'' ACF
(with no approximations) and determined that it could be correctly
described by an ACF with the time dependence obtained in the fast
reaction limit (Eq.~(\ref{eq:G3comp})). The fast reaction
approximation is always valid for large enough $\tau$, but, as shown
in \cite{Ipina2013} for the case of `permanently' fluorescent
molecules, it can still provide a good description of the full ACF for
all $\tau$ even if the reaction and diffusion times are of the same
order. Our results also showed that even if the rapid reaction limit
may not hold, fitting the full ACF with an expression of the form
Eq.~(\ref{eq:G3comp}) still provides reasonably good estimates of the
timescales associated to the free diffusion coefficient of the dye and
to the exponentially decaying term. The two effective coefficients
given by Eqs.~(\ref{eq:Def1})~and~(\ref{eq:Def2}) could also be
estimated for certain dye concentrations. The term that corresponds to
the free diffusion of the dye (Eq.~(\ref{eq:GF})) is exact. Thus, we
can always assume that the weight that corresponds to this timescale
is inversely proportional to the total number of dye molecules in the
observation volume.~The other two individual weights, however, can be
incorrectly assessed if the fast reaction approximation is assumed.
The total weight, on the other hand, is always inversely proportional
to the mean number of Ca$^{2+}$-bound dye molecules in the
observation volume. Thus, in our application of the theory to derive
biophysical parameters from the experimental observations we used the
total weight and the weight of the term that corresponds to the free
diffusion of the dye, but not the other two.

We then performed a series of experiments in solution using $Ca^{\rm
2+}$ and Fluo4 High or Low Affinity at various concentrations.
Fitting the ACF with an expression of the form of
Eq.~(\ref{eq:G3comp}) we obtained the correlation times from which we
derived the corresponding diffusion coefficients as functions of the
dye concentration. As shown in Fig.~\ref{fig:D_efectivos}, one of the
coefficients (or, analogously, the correlation time) remained the same
for all the concentrations. According to the theory, this coefficient
is to be associated with the free diffusion coefficient of the dye. We
observed that the value derived for the dye in its High or Low
Affinity version is approximately the same ($D_{F} =(65\pm 7) \mu
m^{{\rm 2}} /s$ and $D_{F} =(89\pm 8) \mu m^{\rm 2} /s$,
respectively). These values are consistent with the value derived in
solution for 10kDa tetramethylrhodamine-dextran (TMR-D, $85\mu m^{\rm
2} /s$)~\cite{Gennerich_Schild2002}. The variation of the other two
coefficients with the dye concentration is particularly visible in the
case of the High Affinity version of the dye
(Fig.~\ref{fig:D_efectivos}~(a)).

\begin{figure}[!ht]
\begin{center}
\includegraphics{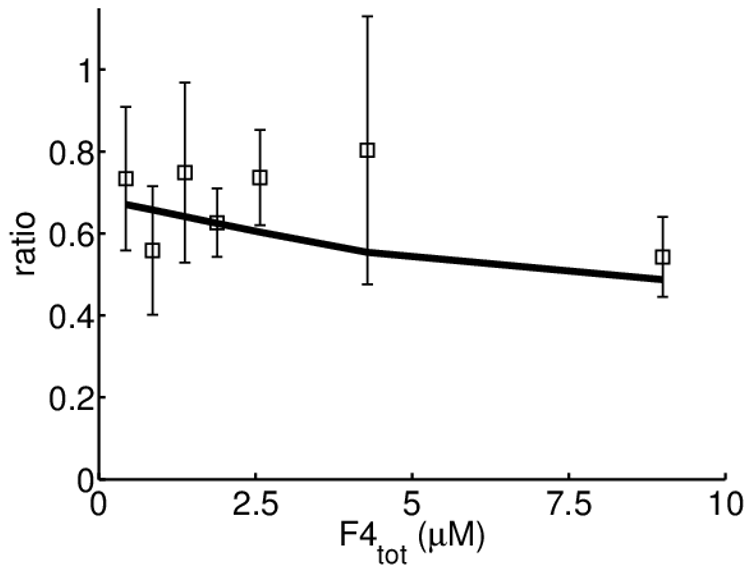}
\end{center}
\caption {\label{fig:Gotot_CaF4}} ~Ratio between the total weights,
$Go_{tot}$ (circles), obtained in experiments with High and Low
Affinity Fluo4 and ratio $CaF_{eq}$(Low)/ $CaF_{eq}$(High), solid
line, determined theoretically as functions of the total dye
concentration, $F4_{tot}$. The $Ca^{{\rm 2+}}$-bound dye
concentrations were computed using the dissociation constant provided
by the vendor.
\end{figure}

We then performed a series of self-consistency checks of our approach.
We first compared the relationship between the inverse of the weights,
$Go_{tot}$ and $Go_{F}$, that we obtained from the experimental fits
and the total concentrations of Fluo4 and of $Ca^{{\rm 2+}}$-bound dye
that we used in the solutions with the theoretical expression,
Eqs.~(\ref{eq:Gotot}) and (\ref{eq:GoF}). The results are shown in
Figs.~\ref{fig:GoF_Gotot_nu}~(a) and~\ref{fig:GoF_Gotot_nu}~(d) with
symbols for the former and curves for the latter. We can observe in
Figs.~\ref{fig:GoF_Gotot_nu}~(a) and~\ref{fig:GoF_Gotot_nu}~(d) that,
in the case of the inverse of $Go_{F}$ versus $F4_{tot}$ relationship,
the experimental points match the theoretical prediction. Thus, for
these experiments in intact cells we expect to be able to obtain a
reliable estimation of the amount of indicator that enters the system.
We fitted the experimental points by a linear relationship between the
inverse of $Go_{F}$ and $F4_{tot}$. We obtained (0.54 ${\rm \pm }$
0.08) $\mu m^{{\rm 3}}$ for High Affinity and (0.56 ${\rm \pm }$ 0.08)
$\mu m^{{\rm 3}}$ for Low Affinity Fluo4. We can observe in
Figs.~\ref{fig:GoF_Gotot_nu}~(b) and~\ref{fig:GoF_Gotot_nu}~(e) that,
in the case of the inverse of $Go_{tot}$ versus $CaF_{eq}$
relationship, the experimental points lie below the theoretical
prediction, as if the actual concentrations of $Ca^{{\rm 2+}}$-bound
dye were smaller than those that can be derived from
Eq.~(\ref{eq:CaF_eq}) using the ones of the solutions and dissociation
constant provided by the vendor. If, as before, we fit the
experimental points by a relationship between the inverse of
$Go_{tot}$ and $CaF_{eq}$ we obtain (0.23 ${\rm \pm }$ 0.02) $\mu
m^{\rm 3}$ for High Affinity and (0.17 ${\rm \pm }$ 0.01) $\mu
m^{\rm 3}$ for Low Affinity Fluo4. The resulting volumes are smaller
than the one determined from the calibration, ($V_{ef} {\rm =}$ (0.59
${\rm \pm }$ 0.1) $\mu m^{{\rm 3}}$), and the mismatch is slightly
larger in the case of Low Affinity Fluo4. We must point out that this
relationship also depends on the dissociation constant of the
$Ca^{\rm 2+}$-dye reaction and that using larger $K_{d}$ values
would decrease the mismatch between the experimental points and the
theoretical curve. In order to analyze to what
extent the results obtained for both dyes agree with what can be
expected theoretically we show in Fig.~\ref{fig:Gotot_CaF4} the ratio
of total weights obtained using each dye (weight for High over weight
for Low Affinity Fluo4 with symbols) as a function of the total dye
concentration for which we had experiments performed with both dyes.
We also show in the figure the ratio of $Ca^{\rm 2+}$-bound dye
concentrations (Low over High) computed theoretically using the
dissociation constant provided by the vendor. These two ratios should
be equal according to Eq.~(\ref{eq:Gotot}). We observe that the ratio
determined experimentally is larger than the theoretical one in most
cases. This implies that either the experimentally estimated value of
$CaF_{eq}$ is underestimated for the High Affinity dye or it is
overestimated for the Low Affinity one. We must recall that
Eq.~(\ref{eq:Gotot}) holds provided that the fluorescence coming from
the free dye molecules is negligible with respect to the one that
comes from the $Ca^{{\rm 2+}}$-bound molecules. Assuming that
Eq.~(\ref{eq:Gotot}) holds in a case in which the free dye molecules
contribution to the fluorescence is not completely negligible would
lead to an overestimation of $CaF_{eq}$. In such a case the
overestimation of $CaF_{eq}$ would be larger for the Low Affinity than
for the High Affinity dye. This could explain the difference between
the experimental points and the theoretical curve of
Fig.~\ref{fig:Gotot_CaF4}. This observation together with the fact
that the mismatch that can be observed in Fig.~\ref{fig:GoF_Gotot_nu}
is larger for the Low than for the High Affinity Fluo4 makes the
latter preferable over the former to study $Ca^{\rm 2+}$ transport
in other settings. Finally, we also compared the dye concentration
dependence of the inverse of the exponential correlation time derived
from the experiments ($\nu$) with the one predicted from the theory
($\nu _{F}$ in Eq.~(\ref{eq:Gef2})) using some estimated parameters as
explained before. As expected from the analyses of
Fig.~\ref{fig:GoF_Gotot_nu}, it is for the lowest dye concentrations
that we obtained comparable results between theory and experiments.

After having tested the self-consistency of our model, we subsequently
used it to derive estimates of some biophysical parameters from the
parameters of the fitting. More specifically, we obtained the free
$Ca^{{\rm 2+}}$ diffusion coefficient, $D_{Ca}$, and the off-rate of
the $Ca^{{\rm 2+}}$-dye binding reaction, $k_{off}$. For the former we
used the sum of the two effective diffusion coefficients ($D_{ef1}$ +
$D_{ef2}$) and subtracted the estimate of the free dye diffusion
coefficient, $D_{F}$. The values, $D_{Ca}$ and $D_{F}$ obtained for
each solution probed are shown in Fig.~\ref{fig:DCa_DF_koff}. The
corresponding average values are within the expected range ($D_{Ca}
{\rm =}$(861 ${\rm \pm }$ 79) $\mu m^{{\rm 2}} {\rm /}s$ , $D_{F}$=
(65 ${\rm \pm }$ 7) $\mu m^{{\rm 2}} {\rm /}s$ in the case of Fluo4
High Affinity and $D_{Ca} {\rm =}$(926 ${\rm \pm }$ 92) $\mu m^{{\rm
2}} {\rm /}s$, $D_{F}$= (89 ${\rm \pm }$ 8) $\mu m^{{\rm 2}} {\rm /}s$
in the case of Fluo4 Low Affinity). In particular, we obtain
consistent values of the free dye diffusion coefficient, $D_{F}$, for
both the High and Low Affinity version (i.e. $D_{F}$ $\sim$ (65-90)
$\mu m^{{\rm 2}} {\rm /}s$) that are of the same order of value as the
one estimated for a 10kDa TMR-D~\cite{Gennerich_Schild2002}. The
estimated free $Ca^{{\rm 2+}}$ diffusion are also consistent with what
we expected~\cite{Qin1991},~\cite{HCP}.

It is important to note that the values, $D_{Ca}$ and $D_{F}$, are
derived exclusively from the diffusive correlation times. Thus, these
results are not affected by the differences between the  $Ca^{\rm
2+}$-bound dye theoretical concentrations, $CaF_{eq}$, and the ones
estimated by the  (\ref{eq:Gotot}) discussed before. In order to
obtain $k_{off}$ from the inverse of the exponential correlation time,
$\nu$, it is necessary to use concentration estimates. In order to
avoid introducing an additional error because of the possible mismatch
between the concentrations that we discussed in connection with the
differences observed in Fig.~\ref{fig:GoF_Gotot_nu}, to obtain
$k_{off}$ from $\nu$ we used the estimates of the ratio between the
concentrations and the dissociation constant derived from the weight,
$Go_{F}$, of the ACF obtained in the experiments. As discussed before,
the values of $\nu$ seemed to display the correct behavior only for
those solutions with the smallest dye concentrations. In any case,
applying the theory to all the experimental results regardless of
$F4_{tot}$ gave values of $k_{off}$ within the same order of magnitude
(see Figs.~\ref{fig:DCa_DF_koff}~(c) and~\ref{fig:DCa_DF_koff}~(e)).
Using the average of these values we obtained $k_{off}$ = (88 ${\rm
\pm }$ 19) s$^{-1}$ and $k_{off}$ = (483${\rm \pm }$ 61) s$^{-1}$
for the High and Low Affinity Fluo4, respectively. Using the
dissociation constant provided by the manufacturer we derived the on
rates. For the High and Low Affinity versions of the dye, we found
similar values ($k_{on}$ = (0.114 ${\rm \pm }$ 0.025) $nM^{-1} s^{-1}$
and $k_{on} = $(0.186 ${\rm \pm }$ 0.023) $nM^{-1} s^{-1}$,
respectively). This is consistent with the fact that, in BAPTA
(1,2-bis(o-aminophenoxy)ethane-N,N,N',N'-tetraacetic acid) based
calcium indicators, increasing $K_{d}$ values results from an increase
in the dissociation rate constant and negligible or only modest
decreases in the association rate
constants~\cite{Tsien1999},~\cite{Naraghi}. It is important to note
that, while concentrations at equilibrium do not depend on $k_{off}$
and $k_{on}$, separately, but on $K_{d}$ = $k_{off}/k_{on}$, their
time evolution does. Therefore, the values of $k_{off}$ and $k_{on}$
affect the behavior of the observed signals and knowing them is
absolutely necessary to infer the spatio-temporal distribution of free
$Ca^{\rm 2+}$ from the images~\cite{Bruno2010},~\cite{Ventura2005}.
Knowing the free diffusion coefficients of $Ca^{\rm 2+}$ and its dyes
in the cytosol is necessary as well for this purpose. The values
derived in the Results Section, however, correspond to coefficients in
aqueous solution. Assuming that the differences in the free diffusion
coefficients in solution and in the cytosol are due to differences in
viscosity between both media we may assume that the ratio between the
free diffusion coefficients of any two substances remains the same in
both settings. Thus, by quantifying the rate of diffusion of a
molecule that diffuses freely in the cytosol and in solution we can
infer the free diffusion coefficient of $Ca^{\rm 2+}$and the dyes in
the cytosol as well. We present in the Appendix the results of FCS
experiments performed in aqueous solution and in oocytes of
\textit{Xenopus laevis} using TMR-D. The ACF can be fitted by an
expression of the form Eq.~(\ref{eq:G1comp}), \textit{i.e}., with a
single, free-diffusing component. From the fits we obtained $D_{TMR}$
=(27 $\pm$ 1) $\mu m^{\rm 2} /s$ in the cytosol considering that the
TMR-D diffusion coefficient in solution is $D_{TMR}$ =85 $\mu m^{{\rm
2}} /s$~\cite{Gennerich_Schild2002}, we obtained
$D_{TMR-D}$(solution)/$D_{TMR-D}$(oocyte) $\sim$ 3.

Assuming that $D_{TMR-D}$(solution)/$D_{TMR-D}$(oocyte) $\sim$
$D_{free}$ (solution)/$D_{free}$(oocyte), where $D_{free}$ stands for
free diffusion coefficient of any substance, we can use the free
transport rates of $Ca^{\rm 2+}$ and of its dyes in solution to infer
their values in the cytosol. We obtain $D_{Ca}$ $\sim$ (261-313)~$\mu
m^{{\rm 2}}/s$ and $D_{F}$ $\sim$ (19-24)~$\mu m^{{\rm 2}}/s$ starting
from the free diffusion coefficients in solution obtained in the
experiments performed with High Affinity Fluo4. Thus, the practical
implementation presented in this paper not only highlights the
advantages of our approach but also allows us to derive information
that is key to quantify the free $Ca^{\rm 2+}$ distribution that
underlies a $Ca^{\rm 2+}$ image.

The cytosolic $D_{Ca}$ values derived with our approach are of the
same order of magnitude as the one obtained in cytosolic extracts by
Allbritton et al., 1992~\cite{Allbritton1992} although the latter
(220~ $\mu m^{\rm 2}/s$) is below our lower bound. The analysis of
buffered diffusion of Pando et al., 2006~\cite{Pando2006}, showed that
the effective diffusion coefficient obtained in the experiments of
Allbritton et al, 1992~\cite{Allbritton1992} is the single molecule
one and a misinterpretation of its meaning could lead to an
underestimation of the actual diffusion rate of $Ca^{\rm 2+}$. This
highlights the need of having an underlying biophysical model to
interpret transport rates in experiments that do not probe solely free
diffusion~\cite{Sigaut2014}. The theory and experiments of this paper
illustrates this very important point. It also shows how by changing
the experimental conditions so that the correlation times associated
to effective diffusion change it is possible to identify the latter
and quantify concentration-independent biophysical parameters. Other
experimentally accessible parameters such as the observation volume
can be modified to change some of the correlation times and, in this
way, quantify different biophysical parameters~\cite{Sigaut2014}. In
fact, a comparison of FCS results obtained for different observation
volumes has recently been used to quantify the binding rates of
transcription factors in single cells of developing mouse
embryos~\cite{White2016}. This shows the relevance of performing FCS
experiments under different conditions to quantify parameters. The
approach presented in this paper can then be extended to address the
quantification of transport rates in other biologically relevant
reaction-diffusion systems.

\section*{Acknowledgments}

We are thankful to Emiliano Perez Ipi\~{n}a for having provided the
code to compute the full ACF and to Lucia Lopez and Estefania Piegari
for help with some of experiments. This research has been supported by
UBA (UBACyT 20020130100480BA) and ANPCyT (PICT 2013-1301). L.S.
and S.P.D. are members of Carrera del Investigador
Cient\'{i}fico (CONICET). 

\section{Appendix}
\subsection{FCS Theory}

\subsubsection{ACF for a system with freely diffusing particles.}

When the fluorescence comes solely from a single type of particles,
$P_{f}$, that diffuse freely with coefficient, $D_{f}$, the
fluorescence is given by: 

\begin{equation}
\label{eq:app_fluorescence}
f(t)=\int QI(r)[P_{f} ](r,t)d^{{\rm 3}} r 
\end{equation}

where $[P_{f} ](r,t)$ is the particle concentration at time, $t$, and
spatial point, $r$, the parameter, $Q$, takes into account the
detection efficiency, the fluorescence quantum yield and the
absorption cross-section at the wavelength of excitation of the
fluorescence. The illumination is commonly approximated by a
three-dimensional Gaussian:

\begin{equation}
\label{eq:app_illumination}
I(r)=I({\rm 0)}\; e^{-\frac{{\rm 2}r^{{\rm 2}} }{w_{r}^{{\rm 2}} } } \; e^{-\frac{{\rm 2}z^{{\rm 2}} }{w_{z}^{{\rm 2}} } } {\rm ,}
\end{equation}

with $z$ the spatial coordinate along the beam propagation direction,
$r$ a radial coordinate in the perpendicular plane and $w_{z} $ and
$w_{r} $ the sizes of the beam waist along $z$ and $r$, respectively.
In this case there is an analytic expression for the ACF which is
given by Eq.~(\ref{eq:G1comp}). Fitting the ACF obtained from
experiments by Eq.~(\ref{eq:G1comp}) two parameters can be determined:
$Go$ and the characteristic time $\tau _{f} $. A previous calibration
of the geometric parameters of the sample volume is required in order
to obtain $D_{f} $ from $\tau _{f} $. This is done performing the same
experiments on a sample for which $D_{f} $ is already known. Once
$w_{r} $ and $w_{z} $ are determined, the unknown $D_{f} $ can be
estimated from the characteristic time $\tau _{f} $ and $P_{tot} $
from $Go$.

\subsubsection{``Full'' ACF of a system with $Ca^{\rm 2+}$ and a single wavelength dye.}

The equations that describe the dynamics of $Ca^{\rm 2+}$ and a single
wavelength dye, $F$, that react and diffuse as described in Sec. II
are:

\begin{equation}
\label{eq:app_dCa_dt}
\frac{\partial [Ca]}{\partial t} {\rm =}D_{Ca} \nabla ^{{\rm 2}} [Ca]-k_{on} [Ca][F]+k_{off} [CaF]
\end{equation}

\begin{equation}
\label{eq:app_dCaF_dt}
\frac{\partial [CaF]}{\partial t} {\rm =}D_{F} \nabla ^{{\rm 2}} [CaF]+k_{on} [Ca][F]-k_{off} [CaF]
\end{equation}

\begin{equation}
\label{eq:app_dF_dt}
\frac{\partial [F]}{\partial t} {\rm =}D_{F} \nabla ^{{\rm 2}} [F]-k_{on} [Ca][F]+k_{off} [CaF]
\end{equation}

In FCS experiments in aqueous solution containing $Ca^{\rm 2+}$ and
$F$ it is assumed that both species uniformly are distributed and in
equilibrium, so that their mean concentrations are given by the
equilibrium concentrations $Ca_{eq}$, $F_{eq}$ and $CaF_{eq}$, that
satisfy Eq.~(\ref{eq:CaF_eq}) and:

\begin{equation}
\label{eq:app_CaF_eq}
CaF_{eq} =\frac{Ca_{eq} F_{tot} }{Ca_{eq} +K_{d} } 
\end{equation}

\begin{equation}
\label{eq:app_Ca_eq}
Ca_{eq} =\frac{1}{2} \left(Ca_{tot} -K_{d} -F_{tot} +\left(\left(Ca_{tot} -K_{d} -F_{tot} \right)^{2} +4K_{d} Ca_{tot} \right)^{1/2} \right)
\end{equation}

\begin{equation}
\label{eq:app_CaF_eq_2}
CaF_{eq} =\frac{1}{2} \left(Ca_{tot} +K_{d} +F_{tot} -\left(\left(Ca_{tot} -K_{d} -F_{tot} \right)^{2} +4K_{d} Ca_{tot} \right)^{1/2} \right)
\end{equation}

In the case in which the calcium indicator is practically
non-fluorescent while it is not bound to $Ca^{\rm 2+}$ the
fluorescence intensity is given by:

\begin{equation}
\label{eq:app_fluo_intensity}
f(t)=\int QI(r)[CaF](r,t)d^{{\rm 3}} r ,
\end{equation}

with $Q$ and $I$ as before. 

As done in Sigaut et al. 2010~\cite{Sigaut2010}, we follow Krischevsky
and Bonnet 2002~\cite{krichevsky2002fluorescence} to determine the
spatio-temporal dependence of the fluorescence fluctuations in this
case. Namely, the evolution
equations~(\ref{eq:app_dCa_dt})-(\ref{eq:app_dF_dt}) are linearized
around the equilibrium solution, Eq.~(\ref{eq:CaF_eq}). The solution
of these linearized equations is then computed in Fourier space and
written in terms of branches of eigenvalues, $\lambda (q)$, and
eigenvectors, $\chi (q)$, with $q$ the variable in Fourier space
(conjugate to the spatial vector $(r,z)$).  The fluorescence
fluctuations are then obtained as in Eq.~(\ref{eq:app_fluo_intensity})
but replacing $[CaF]$ by the corresponding component of the solution
of the linearized problem, $\delta [CaF]$. The calculation of the ACF
finally assumes that the correlation length of the concentrations at
any given time is much smaller than the inter-molecule distance and
that the number of molecules obeys Poisson statistics so that its
variance and its mean are equal. In this way the ACF, $G(\tau )$, can
be written as a sum of as many components as branches of eigenvalues
of the linearized problem, in this case:

\begin{equation}
\label{eq:app_G_tau}
G(\tau )=G_{\lambda _{F} } (\tau )+G_{\lambda _{1} } (\tau )+G_{\lambda _{2} } (\tau)
\end{equation}

with:

\begin{equation}
\label{eq:app_G_lambda_F}
G_{\lambda _{F} } (\tau )=\frac{Go_{F} }{\left({\rm 1}+\frac{\tau }{\tau _{F} } \right)\sqrt{{\rm 1}+\frac{\tau }{w^{{\rm 2}} \tau _{F} } } } 
\end{equation}

\begin{equation}
\label{eq:app_G_lambda_1}
G_{\lambda _{1} } (\tau )=\frac{1}{2(2\pi )^{3} hCaF_{eq} } \int d^{3}  qI(q)\left(1+\frac{(a-h)\nu _{F} }{(a+h)\Psi (q)} +\frac{(D_{Ca} -D_{F} )q^{2} }{\Psi (q)} \right)e^{\lambda _{1} t}  
\end{equation}

\begin{equation}
\label{eq:app_G_lambda_2}
G_{\lambda _{2} } (\tau )=\frac{1}{2(2\pi )^{3} hCaF_{eq} } \int d^{3}  qI(q)\left(-1-\frac{(a-h)\nu _{F} }{(a+h)\Psi (q)} +\frac{(D_{Ca} -D_{F} )q^{2} }{\Psi (q)} \right)e^{\lambda _{2} t}
\end{equation}

where $I(q)=\exp (-(w_{_{r} }^{2} q_{_{r} }^{2} +w_{z}^{2} q_{_{z}
}^{2} )/4)$, $q_{r} $ and $q_{z} $ the Fourier coordinates conjugated
to the radial and axial coordinates, $r$ and $z$, respectively,
$a=F_{eq} /K_{d} $, $h=F_{T} /F_{eq} $, $\nu _{F} =k_{off} (a+h)$,
$\Psi (q)=\sqrt{(D_{F} -D_{Ca} )^{2} q^{4} +2q^{2} (D_{F} -D_{Ca}
)(h-a)k_{off} +(h+a)^{2} k_{off}^{2} } $ and the eigenvalues:

\begin{equation} 
\label{eq:app_lambda_1}
\lambda _{1} =-\frac{1}{2} \left(k_{off} (a+h)+\left(D_{F} +D_{Ca} \right)q^{2} \right)+\frac{\Psi }{2} 
\end{equation}

\begin{equation}   
\label{eq:app_lambda_2}
\lambda _{2} =-\frac{1}{2} \left(k_{off} (a+h)+\left(D_{F} +D_{Ca} \right)q^{2} \right)-\frac{\Psi }{2} 
\end{equation}

\subsubsection{Approximated ACF of a system with $Ca^{{\rm 2+}} $ and a single wavelength dye.}

Although $G_{\lambda _1} (\tau )$ and $G_{\lambda _2} (\tau )$ can be
computed numerically, in general there is no analytic algebraic
expression for these two components as there is for the one that
corresponds to the branch of eigenvalues, $\lambda _{F} =-D_{F}
q^{2}$, associated to the free diffusion coefficient of the dye,
$D_{F}$ (see Eq.~(\ref{eq:app_G_lambda_F})). As done in Sigaut et al.
2010~\cite{Sigaut2010}, however, an analytic expression for
$G_{\lambda _1} (\tau)$ and $G_{\lambda _2} (\tau)$ and, consequently,
for the ACF can be obtained in the limit of small $q$ which is always
valid for long enough times, $\tau$. The approximation is good for
almost any value of $\tau$ when the observation volume is such that
the characteristic reaction time is of the same order or less than the
diffusive time across the volume~\cite{Ipina2013}. In fact, if we
expand the integrands that define $G_{2} (\tau)$ and $G_3 (\tau)$ in
powers of $q$ and keep the expansion up to $O(q^{\rm 2})$ we obtain
Eq.~(\ref{eq:Gapprox}) This limit is valid provided that the reactions
occur on a faster timescale than diffusion across the observation
volume, i.e., if $\tau _{reac} \equiv \left(k_{off} +k_{on} (Ca_{eq}
+F_{eq} )\right)^{-1} <<w_{r} ^{2} /(4D_{Ca} )$.

\subsection {Limits of applicability of the fast reaction approximation}

\begin{figure}[!ht]
\begin{center}
\includegraphics {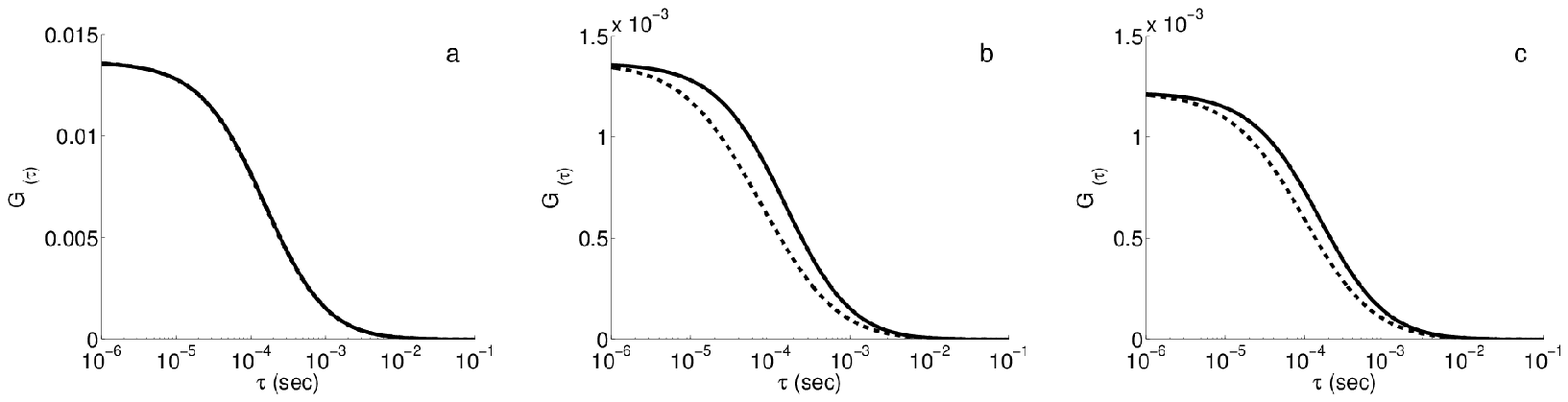}
\caption {\label{fig:ACF_sol}} {Full (solid line,
Eqs.~(\ref{eq:app_G_tau})-(\ref{eq:app_G_lambda_2})) and approximated 
(dashed line, Eqs.~(\ref{eq:Gapprox})-(\ref{eq:Gef2})) ACF's for Fluo4
High Affinity using the parameters listed in
Table~\ref{tab:app_parametros}. $F4_{tot}$ = 429~nM (a), 7500~nM (b),
15~$\mu M$ (c).}
\end{center}
\end{figure}

In order to study when the fast reaction approximation of the ACF can
be used to estimate different biophysical parameters we computed
numerically the full ACF, $G(\tau)$,given by
Eqs.~(\ref{eq:app_G_tau})-(\ref{eq:app_G_lambda_2}) using an adaptive
Lobatto quadrature algorithm, with the\textit{quadl} function on the
MatLab platform (The MathWorks, Natick, MA) and the parameters listed
in Table~\ref{tab:app_parametros}. We compared the results of these
computations with the approximated ACF, $G_{approx} (\tau)$, given by
Eqs.~(\ref{eq:Gapprox})~-~(\ref{eq:Gef2}) using the same parameters.
For the comparison we computed the difference between both functions
given by:

\begin{equation}
\label{eq:app_epsilon}
\varepsilon ^{2} =\frac{1}{n} \sum _{i=1}^{n}\left(G(\tau _{i} )-G_{approx} (\tau _{i} )\right)^{2} 
\end{equation}

with \textit{n} the total number of data points. For the lowest dye
concentrations considered $G(\tau)$ and $G_{approx}(\tau)$ were
indistinguishable. As the concentration of dye was increased, the
difference between the full and the approximated ACF's first
increased, with $G_{approx}(\tau)$ decaying at an earlier correlation
time than $G(\tau)$. The difference between $G(\tau)$ and
$G_{approx}(\tau)$ reached a maximum at $F4_{tot}$ $\sim$ 4 $\mu M$.
Further increments in $F4_{tot}$ decreased this difference. This is
illustrated in Fig.~\ref{fig:ACF_sol} where we show $G(\tau)$ and
$G_{approx}(\tau)$ with solid and dashed lines, respectively, for
$F4_{tot}$=429 $nM$, 7500 $nM$, 15 $\mu M$ using the parameters of
Fluo4 High Affinity. Similar results are obtained for Fluo4 Low
Affinity (data not shown). The difference between the two ACF's,
however, is never significantly large: we obtained 2.31x10 $^{-9}$
$\le$ $\varepsilon ^{2}$ ${\le}$ 1.17x10$^{-8}$ for High Affinity
Fluo4 and 1.55x10 $^{-8}$ ${\le}$  $\varepsilon ^{2}$ ${\le}$ 4.11x10
$^{-8}$ for Low Affinity Fluo4. The differences between the individual
components associated to $\tau _{ef1}$ and $\tau _{ef2}$ are much
larger. 

\begin{table}[!t]
\centering
\caption {\label{tab:app_parametros} Parameters used to compute the full and
 approximated ACFs numerically. For the concentrations of dye we tried the 
 values listed in Table \ref{tab:soluciones_F4_Ca}.}
\begin{tabular}{lll}
\hline\hline
Parameter  & \multicolumn{2}{c}{Value}\\ 
$w_{r}$ & \multicolumn{2}{c}{0.28 $\mu m$}\\
$w$  & \multicolumn{2}{c}{5}\\
$D_{Ca}$& \multicolumn{2}{c}{760  $\mu m^{\rm 2} {\rm /}s$}\\
$D_{F}$ & \multicolumn{2}{c}{85 $\mu m^{\rm 2} {\rm /}s$}\\
$Ca_{tot}$ & \multicolumn{2}{c}{4285 $nM$}\\
 & \multicolumn {1} {c} {High Affinity Fluo4} & \multicolumn {1} {c} {Low Affinity Fluo4}\\
$K_{d}$  &\multicolumn {1} {c} {772 $nM$} & \multicolumn {1} {c} {2600 $nM$}\\
$k_{off}$  & \multicolumn {1} {c} {80$s^{-1}$} & \multicolumn {1} {c} {300 $s^{-1}$}\\
\hline\hline 
\end{tabular}
\end{table}

\begin{figure}[!ht]
\begin{center}
\includegraphics{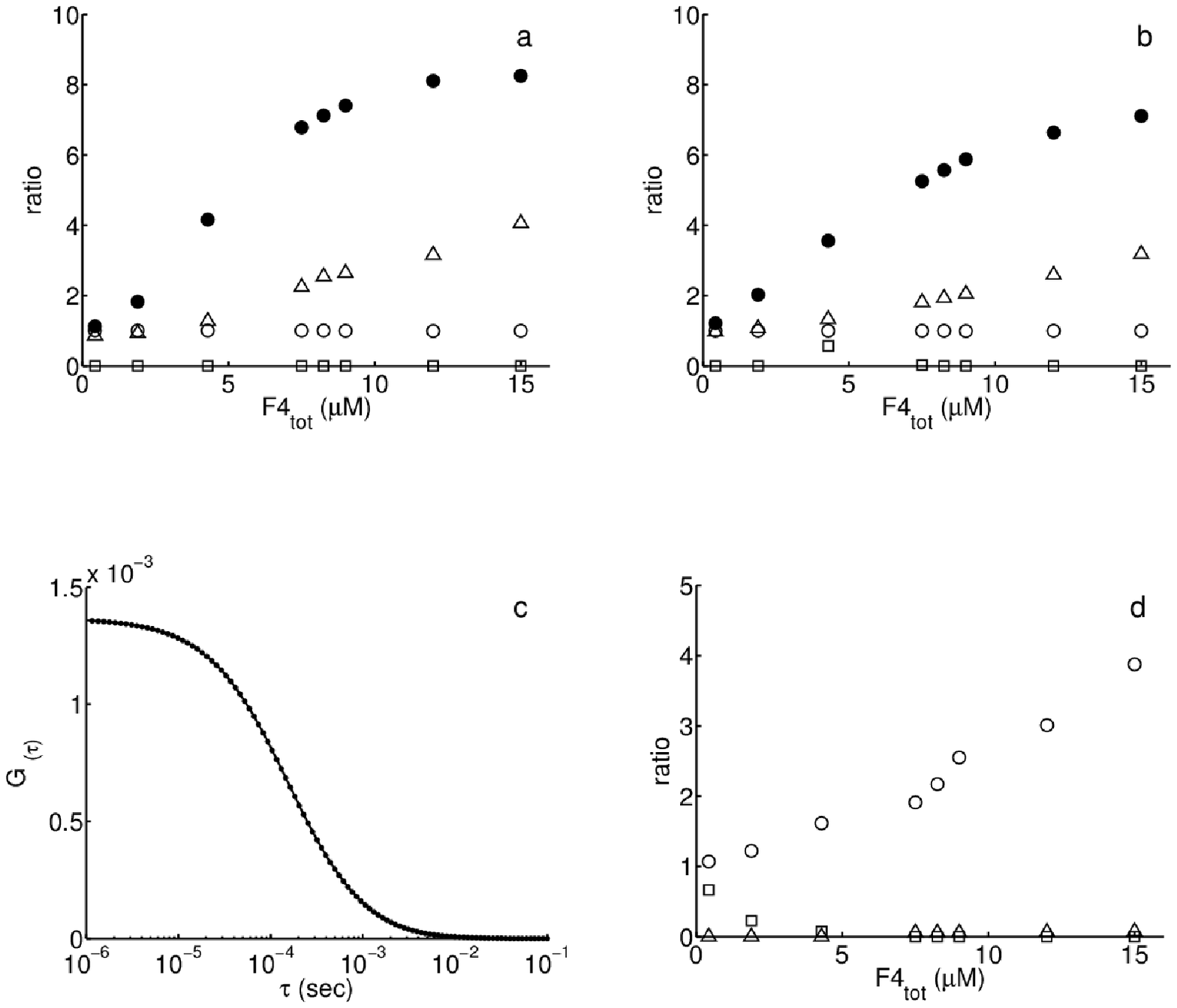}
\caption {\label{fig:ratios}} {Ratios $\tau _{0} /\tau _{F}$ (white
circles), $\tau _{2} /\tau _{ef2}$ (black circles), $\nu _{F} /\nu$
(triangles) between the fitted values and those of the fast reaction
approximation and $G_{1} {\rm /}G_{0}$ (squares), as a function of
$F4_{tot}$ for High Affinity (a) and Low Affinity (b) Fluo4. (c) Full
ACF (dotted line) fitted with Eq.~(\ref{eq:G3comp}), fixing the
timescales (solid line) for $F4_{tot}$=7500nM. (d) Ratios between the
fitted weights with the timescales fixed and the fast reaction
approximation weights, $G_{0}/Go_{F}$ (circles), $G_{1}/Go_{ef1}$
(triangles), $Go_{2}/Go_{ef2}$ (squares), as a function of $F4_{tot}$
for Fluo4 High Affinity.}
\end{center}
\end{figure}

We then analyzed what correlation times could be derived by fitting
the full ACF with Eq.~(\ref{eq:G3comp}). We probed two options. First,
we fixed the timescales as in the fast reaction approximation and
fitted the weights. Secondly, we fitted both the weights and the
timescales. From the second test we determined that the fitted values
obtained for $\tau _{0}$ were similar to those prescribed by the fast
reaction approximation, $\tau _{F}$, for all dye concentrations. The
fitted values of $1/\nu $ were similar to the values of the fast
reaction approximation for dye concentrations below 4.826~$\mu M$. For
higher dye concentrations the fitted values of $1/\nu $ for High
Affinity Fluo4 followed the same pattern and stayed within the same
order of magnitude as the value of the fast reaction approximation
although it got three times the approximated value at
$F4_{tot}$ = 12~$\mu M$. For Low Affinity Fluo4 the variations of 
$1/\nu$ with $F4_{tot}$ were slightly different but $1/\nu$ stayed
within the fast reaction approximation values for all dye
concentrations becoming between twice and three times smaller at
$F4_{tot}$ = 15~$\mu M$ For all dye concentrations we obtained 
$\tau _{0}=\tau _{F}$ and $G_{1} <<G_{0}$ and for dye concentration
below 8.25~$\mu M$ we obtained $\tau _{2} \approx \tau _{0}$. These
results are illustrated in Figs.~\ref{fig:ratios}~(a)~-~(b) where we
show the ratios $\tau _{0} /\tau _{F}$, $\tau _{2} /\tau _{ef2}$, 
$\nu _{F}/\nu$, between the fitted values and those of the fast reaction
approximation and $G_{1} {\rm /}G_{0} $, as a function of $F4_{tot}$
for High Affinity (Fig.~\ref{fig:ratios}~(a)) and Low Affinity
(Fig.~\ref{fig:ratios}~(b)) Fluo4. From the test we determined that
the full ACF could be approximated fairly well using the expression
given by Eq.~(\ref{eq:G3comp}), with the timescales of the fast
reaction approximation but with slightly different weights. This is
illustrated in Fig.~\ref{fig:ratios}~(c) where we have plotted these
two ACF's for High Affinity Fluo4 at $F4_{tot}$ = 7500~nM. Similar
figures are obtained for Low Affinity Fluo4 and at other dye
concentrations (data not shown). In this case the mismatch,
$\varepsilon$ $^{2}$, obtained ranged between 7.2x10$^{-12}$ and
1.57x10$^{-10}$ for High Affinity and between 9.69x10$^{-11}$ and
2.46x10$^{-9}$ for Low Affinity Fluo4. Regarding the individual
components of the fitted ACF, the weights obtained, $G_{0}$, $G_{2}$,
were of the same order of magnitude as those of the fast reaction
approximation, $Go_{F}$, $Go_{ef{\rm 2}} $, and $G_{1} $was negligible
for low dye concentrations. This is illustrated in
Fig.~\ref{fig:ratios}~(d) where we have plotted the ratios between the
fitted and the fast reaction approximation weights, $G_{0}/Go_{F}$,
$G_{1}/Go_{ef1}$, $Go_{2}/Go_{ef2}$, as a function of $F4_{tot}$ for
Fluo4 High Affinity. Similar patterns are observed for Fluo4 Low
Affinity (data not shown). 

\subsection{FCS experiments in aqueous solution and in \textit{Xenopus laevis} oocytes with tetramethylrhodamine-dextran to determine the factor by which free diffusion coefficients are rescaled in the cytoplasm.}

We here present the results of performing FCS experiments with
tetramethylrhodamine-dextran (TMR-D) in aqueous solution and in
\textit{Xenopus laevis} oocytes. The aim of these experiments is to
determine the conversion factor between free diffusion coefficients in
the two media. 

\textit{X. laevis} oocytes, previously treated with collagenase and
stored in Barth's solution, were loaded with 37 $nl$ of TMR-D at
different concentrations. Intracellular microinjections were performed
using a Drummond microinjector. Assuming a 1~$\mu l$ cytosolic volume,
the final concentration of TMR-D was 0.9, 1.1, 1.4 or 1.85 $\mu M$.
FCS measurements were performed on a spectral confocal scanning
microscope FluoView 1000 (Olympus, Tokyo, Japan), employing a 60x,
1.35 N.A. oil-immersion objective (UPlanSAPO, Olympus) and a pinhole
aperture of 115~$\mu m$. Single point measurements at a 50~kHz sampling
rate were performed for a total duration of 167~s (equivalently,
8365312 data points) employing a 543~nm line and detecting the
fluorescence in the range (555-655)~nm. For the aqueous solutions the
measurements was performed at $\sim$20~$\mu m$ from the coverslip and
for the oocytes, at the cortical granules region in the animal
hemisphere. Experimental ACF's were calculated with a custom-made
routine written on the Matlab platform~\cite{MATLAB}. To this end,
each 167~s long record was divided into \textit{N$_{sol}$}=1021, 164~ms
long segments containing 2${}^{13}$ points each for the experiments in
aqueous solutions and into \textit{N$_{oo}$}=510, 328~ms long segments
containing 2$^{14}$ points each for the experiments in \textit{X.
laevis} oocytes. The ACF was computed for each of the
\textit{N$_{sol}$}=1021 or \textit{N$_{oo}$}=510 segments from which
the average ACF was obtained. As the confocal volume dimensions are
wavelength-dependent we used the FCS experiments with TMR-D in
solution to estimate the beam waist and aspect ratio at 543~nm.
Assuming a diffusion coefficient of $D_{TMR-D} = 85~\mu m^{\rm
2}/s$~\cite{Gennerich_Schild2002} we obtained $w_r = (0.199\pm 0.003)~
\mu m$ and $w=wz/wr=5$. The ACF was fitted using only one (diffusive)
component as in Eq.~(\ref{eq:G1comp}).

\begin{figure}[!ht]
\begin{center}
\includegraphics{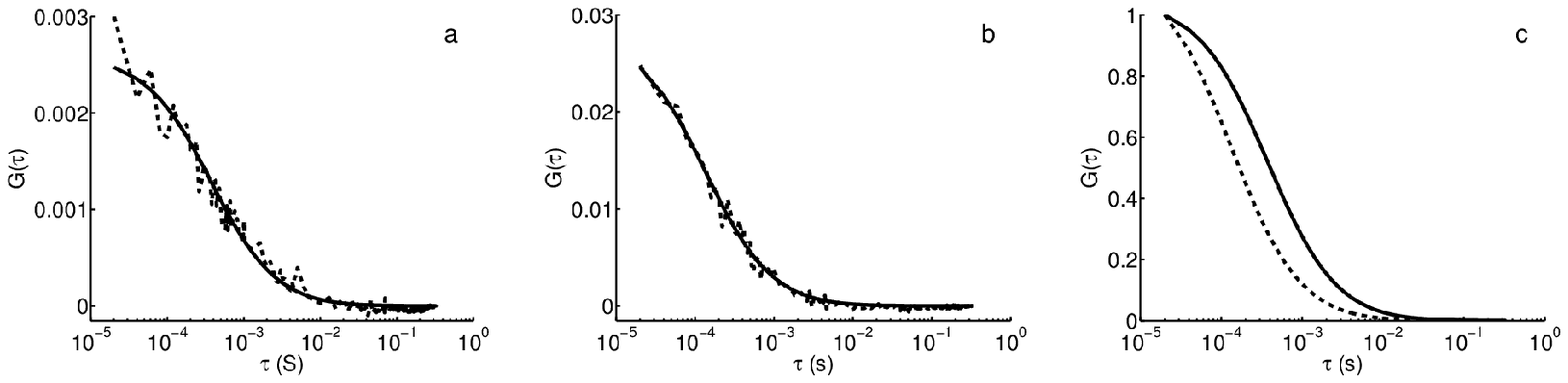}
\caption {\label{fig:sol_ovo}} (a) ACF obtained from FCS experiments
performed in \textit{X. laevis} oocytes microinjected with 37~$nl$ of
TMR-D~=~30 $\mu M$ (dashed line) fitted by Eq.~(\ref{eq:G1comp}) (solid
line). (b) As in (a) for solution of TMR-D =50nM. (c) ACF's from the
fits performed in (a) and (b) (solid and dashed line, respectively), 
normalized. 
\end{center}
\end{figure}

We show in Fig.~\ref{fig:sol_ovo} the ACF obtained from FCS
experiments performed in \textit{X. laevis} oocytes with TMR-D
(Fig.~\ref{fig:sol_ovo}~(a)). Using Eq.~(\ref{eq:G1comp}) to fit the
data of Fig.~\ref{fig:sol_ovo} we obtain $D_{TMR}(oocyte)=(27\pm 1)~\mu
m^{{\rm 2}} /s$. The TMR-D diffusion coefficient in solution is
$D_{TMR}(solution)=85~\mu m^{\rm 2}/s$~\cite{Gennerich_Schild2002}.
Thus, it is $D_{TMR-D}$(oocyte)/$D_{TMR-D}$(solution) $\sim$ 3. 

\end{document}